\documentclass[preprint,amssymb,aps,prb,showpacs,showkeys]{revtex4-1}

\usepackage[utf8x]{inputenc}
\usepackage[T1]{fontenc}
\usepackage{amsmath}
\usepackage{color}
\usepackage{graphicx}
\DeclareMathOperator\erfc{erfc}

\begin{document}
\date{\today}
\title{Tuning the critical magnetic field of the triplon Bose-Einstein condensation in \(\boldsymbol{\mathbf{\mathrm{Ba}_{3-x}\mathrm{Sr}_x\mathrm{Cr_2O_8}}}\)}
\author{Henrik Grundmann}
\email{grundmann@physik.uzh.ch}
\author{Alsu Gazizulina}
\author{Andreas Schilling}
\affiliation{Physik-Institut, Universit\"at Z\"urich, Winterthurerstrasse 190, CH-8057 Z\"urich, Switzerland}
\author{Fabian von Rohr}
\affiliation{Department of Chemistry, Princeton University, Princeton, NJ 08544, USA}
\author{Tobias Förster}
\affiliation{Hochfeld-Magnetlabor Dresden (HLD-EMFL), Helmholtz-Zentrum Dresden-Rossendorf, D-01314 Dresden, Germany}
\author{Laurens Peters}
\affiliation{High Field Magnet Laboratory (HFML - EMFL), Radboud University, Toernooiveld 7, 6525 ED NIJMEGEN, The Netherlands}

\begin{abstract}
\noindent 
The structure and magnetic interactions of the triplon Bose-Einstein condensation candidates Ba\(_3\)Cr\(_2\)O\(_8\) and Sr\(_3\)Cr\(_2\)O\(_8\) have been studied thoroughly in the literature, but little is known about a possible triplon condensation in the corresponding solid solution Ba\(_{3-x}\)Sr\(_x\)Cr\(_2\)O\(_8\). We have prepared various members of this solid solution and systematically examined their magnetic properties in high magnetic fields up to \(60\,\mathrm{T}\) and at low temperatures down to \(340\,\mathrm{mK}\), by means of pulsed field and cantilever magnetometry. From these experiments for \(x\in\{3,2.9,2.8,2.7,2.6,2.5\}\), we find that the critical fields of Ba\(_{3-x}\)Sr\(_x\)Cr\(_2\)O\(_8\) decrease monotonically with decreasing Sr content \(x\). This change is in good agreement with the earlier reported variation of the magnetic interactions in these compounds.
\end{abstract}
\maketitle

\section{Introduction}
Over the past two decades, spin dimer systems have been in the focus of intense research due to their exotic magnetic properties. Several spin dimer materials have been shown to exhibit a spontaneous increase of the magnetization upon cooling below a certain temperature \(T_c(H)\) in magnetic fields \(H\) beyond a critical magnetic field \(H_c\). This spontaneous increase has been interpreted in terms of the Bose-Einstein condensation (BEC) of triplons.\cite{nikuni_tlcucl3_magnetization} The longitudinal magnetization \(M_z\) (i.e., the component parallel to the external magnetic field, which is associated with the condensate density) increases in such a way as one would expect it from an increase of the number of condensed bosons below \(T_c(H)\) in a typical BEC.\cite{dellamore_tlcucl3_triplon_density}
Alternatively, \(M_z\)  increases, to first approximation, virtually linearly with \(H-H_c(T)\) at a fixed temperature \(T\). First discovered in TlCuCl\(_3\), this effect was found to take place in a number of compounds such as Pb\(_2\)V\(_3\)O\(_9\)\cite{waki_pb2v3o9}, NiCl\(_2\cdot\)4SC(ND\(_2\))\(_2\)\cite{zapf_dtn} (DTN), BaCuSi\(_2\)O\(_6\)\cite{jaime_bacusi2o6}, Ba\(_3\)Cr\(_2\)O\(_8\)\cite{kofu_symmbreak}, and Sr\(_3\)Cr\(_2\)O\(_8\)\cite{aczel_sr3cr2o8}. Especially the latter two materials have been examined in numerous investigations, both targeting their spin systems\cite{kofu_ba3_bec, aczel_ba3cr2o8,aczel_sr3cr2o8,mrbpaper,grundmann_ww-aenderung} and their general electronic structure. \cite{chapon,kofu_symmbreak,whangbo_ba3cr2o8,radtke_sr3cr2o8}
Ba\(_3\)Cr\(_2\)O\(_8\) and Sr\(_3\)Cr\(_2\)O\(_8\) are isostructural with a three-dimensional arrangement of Cr\(^{5+}\) dimers that feature a dominant intradimer interaction constant \(J_0\) and weaker interdimer interactions \(J'\). The crystal structure at room temperature can be described using the highly symmetric space group R\(\overline{3}\)m\cite{chapon,kofu_symmbreak} which leads to magnetic frustration of the spin system. Both materials undergo a Jahn-Teller induced structural phase transition upon cooling that lifts this magnetic frustration and strongly modifies the magnetic interactions in the system, thereby strengthening \(J_0\).\cite{chapon} This phase transition has been shown to be gradually suppressed in the solid solution Ba\(_{3-x}\)Sr\(_x\)Cr\(_2\)O\(_8\) for an intermediate Sr content \(x\).\cite{mrbpaper} As this gradual suppression in turn strongly influences the magnetic interactions, \(J_0\) exhibits a non-monotonous decrease from Sr\(_3\)Cr\(_2\)O\(_8\) to Ba\(_3\)Cr\(_2\)O\(_ 8\) with a minimum for \(J_0\) around \(x\approx2\).\cite{grundmann_ww-aenderung}

The critical magnetic field \(H_{c1}\) in these systems given by the spin gap \(\Delta\),\cite{giamarchi_bec} which in turn depends on the magnetic interaction constants \(J_0\) and \(J'\).\cite{quinterocastro_sr3cr2o8} Thus, a change of \(J_0\) should be accompanied by a corresponding modification of \(H_{c1}\), thereby changing the dome-like phase boundary \(T_c(H)\). However, no study of the critical fields of the solid solution Ba\(_{3-x}\)Sr\(_x\)Cr\(_2\)O\(_8\) has been reported to date. To determine the dependence of the critical field on the Sr content, we examined several members of the solid solution Ba\(_{3-x}\)Sr\(_x\)Cr\(_2\)O\(_8\) with \(x\in\{3,2.9,2.8,2.7,2.6,2.5\}\) at various temperatures. In this work, we describe the corresponding high-field magnetometry experiments and compare the resulting data for \(H_c(T,x)\) with the changes of the interaction constant \(J_0(x)\) and the reported phase diagram for pure Sr\(_3\)Cr\(_2\)O\(_8\).

\section{Experimental details}
\subsection{Synthesis}
The samples were synthesized as polycrystalline powders using standard solid-state reaction schemes. Ba(NO\(_3\))\(_2\), Sr(NO\(_3\))\(_2\) and Cr(NO\(_3\))\(_2\cdot\)9H\(_2\)O were mixed according to 
\begin{align*}
     \text{(3-x)Ba(NO}_3\text{)}_2 + \text{xSr(NO}_3\text{)}_2 + \text{2Cr(NO}_3\text{)}_2\cdot\text{9H}_2\text{O}\rightarrow&\text{Ba}_\text{3-x}\text{Sr}_\text{x}\text{Cr}_2\text{O}_8+\\
&+\text{10NO}_2+\text{O}_2+9\text{H}_2\text{O},
\end{align*}
dissolved in water and heated afterwards while {continuously} stirred to keep the mixture homogeneous. After evaporating the water, the remaining powder was ground and heated under flowing argon at 915\,\(^{\circ}\)C for 24\,h to remove any excess water and NO\(_x\). The resulting oxide powders were ground again, pressed into pellets and sintered at 1100\,\(^\circ\)C for 48\, h under flowing Ar.

\subsection{Pulsed-field magnetometry}
The \(\chi(H)=\frac{\mathrm{d}M}{\mathrm{d}H}\) data were obtained from magnetometry experiments in pulsed fields up to \(\mu_0H_\text{max}=60\,\mathrm{T}\) at the Hochfeld-Magnetlabor Dresden (HLD) of the Helmholtz-Zentrum Dresden-Rossendorf (HZDR). We used a standard \(^4\)He flow cryostat designed for temperatures down to \(T\approx1.5\,\mathrm{K}\). The samples were placed in a Teflon tube with no relevant magnetic background.
To be able to compare our results directly to the heat capacity and magnetocaloric measurements of Aczel \textit{et al.} on Sr\(_3\)Cr\(_2\)O\(_8\)\cite{aczel_sr3cr2o8}, we first performed magnetometry experiments on a polycrystalline sample of pure Sr\(_3\)Cr\(_2\)O\(_8\). As our obtained values for \(H_\mathrm{c}(T)\) agreed well with the published results, we proceeded to examine the susceptibility \(\chi(H)\) for additional samples with \(x\in\{2.9,2.8,2.7,2.6,2.5\}\) at temperatures between \(T=1.4\,\mathrm{K}\) and \(T=10\,\mathrm{K}\).
The magnetic field was determined by time-integration of the voltage \(U_B\propto\frac{\partial B}{\partial t}\) induced by the magnetic field in a pick-up coil and applying a calibration factor provided by the HLD. The sample signal was observed using a set of compensated coils. This resulted in a pick-up voltage \(U_M\propto\frac{\partial M_\text{sample}}{\partial t}\)  without the signal of the pulsed external field. 
The susceptibility was then obtained as the ratio of the two voltage signals \(\chi_\text{raw}=\frac{U_M}{U_B}\). As no calibration of the pick-up signals has been performed, \(\chi_\text{raw}\) is not equal, but only proportional to the true sample susceptibility \(\chi\). However, for reasons of simplicity and as no absolute values of \(M\) and \(\chi\) are used in this work, we refer to \(\chi_\text{raw}\) as \(\chi\) in the following.
Measurements of the empty magnetometer did only yield a smooth, featureless background \(\chi_\text{BG}\). Thus, no background correction was applied to the resulting data for any of the examined samples. For fields above \(40\,\)T, the experimental noise was too large to perform a reliable numerical differentiation. Thus, only data for magnetic fields below \(40\,\)T have been used for our analysis and are shown in this work.

\subsection{Cantilever magnetometry}
The \(M(H)\) data at temperatures below \(T=1\,\mathrm{K}\) were obtained from cantilever magnetometry experiments for samples with \(x=\{2.9,2.8\}\) at the High Field Magnet Laboratory (HFML) in Nijmegen. The used method is based on measuring the change of the capacitance between a reference plate and a BeCu cantilever with the sample attached as a function of the magnetic field. This change is due to a slight bending of the cantilever by the torque exerted on the sample in an external magnetic field. The capacitance was examined using a \textit{Andeen-Hagerling AH 2700A} capacitance bridge and a \textit{Stanford Research SR830} lock-in amplifier and the field was applied perpendicular to the cantilever.
As the samples were polycrystalline and thus very isotropic, no torque can be measured in the center of the field. Accordingly, the samples were placed outside the field center where the field exhibits a certain gradient. To improve the sensitivity at temperatures below \(2\,\mathrm{K}\), the samples were placed between \(1\,\mathrm{cm}\) and \(3\,\mathrm{cm}\) above the field center, depending on the desired maximum field.

Due to the geometry of the experiment, the magnetic moment \(M\) of the sample can be calculated as \(M=A\left(\frac{1}{C(H)}-\frac{1}{C(H=0)}\right)\frac{1}{\mathrm{grad}H(x)}\), where \(C\) is the capacitance of the cantilever setup, \(A\) a constant factor and \(x\) the sample position. As only relative changes of \(M\) are considered in this work, the value of \(A\) was not determined. The strength of the magnetic field and the gradient itself have been calculated based on the sample position \(x\) and the measured current \(I\) through the coils and using calibration curves provided by the HFML. 

During several test measurements, the capacitance change was found to be antisymmetric with respect to the field center when placing the sample below or above. Thus, we are certain that changes of the capacitance signal are only due to variation of the sample magnetization. Similar to the case of the pulsed-field data, no background subtraction was conducted, as the BeCu cantilever itself only gives a negligible signal.

\subsection{Analysis}
As in the case of pure Sr\(_3\)Cr\(_2\)O\(_8\)\cite{aczel_sr3cr2o8}, no hysteresis was found upon reversing the direction of the variation of \(H\) for the triplon phase transition. Thus, our data analysis is based on the assumption that the triplon BEC is a second order phase transition, as expected. This implies a discontinuity in the second derivatives of the Gibbs free enthalpy\cite{nolting}. As the magnetization is the first derivative of the Gibbs enthalpy with respect to the external field, \(M=\frac{\mathrm{d} G}{\mathrm{d} H}\), a step-like feature should be observable in \(\chi(H)=\frac{\mathrm{d}^2G}{\mathrm{d}H^2}=\frac{\mathrm{d}M}{\mathrm{d}H}\) leading to a peak in \(\rho(H)=\frac{\mathrm{d}^2M}{\mathrm{d}H^2}\). The position of the peak in \(\rho(H)\) is usually taken as the critical field \(H_c\) (see below).\cite{sebastian_first,sebastian_second}
The derivatives were numerically obtained as difference quotient with subsequent smoothing through a symmetric running average. The smoothing window was kept smaller than 40\(\%\) of the full width at half maximum of the observed peak in \(\rho(H)\), so that no significant additional broadening was introduced. 

Determining the position of the peak in \(\rho(H)\) from the maximum value of the second numerical derivative of \(M(H)\) did not yield reliable results due to the significant noise level. The thus determined values of the critical field depended strongly on the
chosen smoothing window. We have therefore decided to fit the peak in \(\rho(H)\) using a analytical function and determine \(H_c\) from the maximum of this function. A common choice for fitting the peak in \(\rho(H)\) is a Gaussian function with a symmetric shape.\cite{zheludev} However, it became clear that our \(\rho(H)\) data are always slightly asymmetric with a tail towards lower fields, even for pure Sr\(_3\)Cr\(_3\)O\(_8\) (see, e.g., Fig. \ref{fig:comparison_at_low_T}). This kind of asymmetry is not exclusive to Ba\(_{3-x}\)Sr\(_x\)Cr\(_2\)O\(_8\), but can also be found in the compound
NiCl\(_2\cdot\)4SC(ND\(_2\))\(_2\).\cite{zheludev}. In our system, it becomes much more pronounced at high temperatures and especially for intermediate values of the Sr content \(x\). This temperature dependent asymmetry can be accomodated for by convoluting the Gaussian function by a temperature dependent exponential:
\begin{align}
 \rho^\text{calc}(H)&=\int\limits_{-\infty}^{\infty}e^{\frac{\mu_0(H-H_c)}{\alpha k_BT}}\theta(H_c-H)\frac{\sqrt{2}\mu_0}{\sqrt{\pi}\sigma}e^{-\left(\frac{\mu_0(H_c-H_0)}{\sqrt{2}\sigma}\right)^2}\mathrm{d}H_c\nonumber\\
		    &=Ae^{\frac{\mu_0(H-H_0)}{\alpha k_B T}}\erfc\left(\mu_0\frac{H-H_0+\frac{\sigma^2}{\mu_0\alpha k_B T}}{\sqrt{2}\sigma}\right),\label{eqn:fitfkt}
\end{align}

where \(\erfc(x)\) is the complementary error function. The resulting fit to our data (with free parameters \(H_0(x,T)\), \(\sigma(x,T)\) and \(\alpha(x)\)) is excellent (see Fig. \ref{fig:comparison_at_low_T}) and allows us to reliably determine the maximum of \(\rho(H)\) using numerical methods.

The critical fields \(H_c\) were then taken as this maximum. The uncertainty of \(H_{c}\) was defined by allowing the sum of the squared residuals, \(\sum\limits_i(\rho_i^\text{exp}-\rho_i^\text{calc})^2\) to be twice the optimal value.

\section{Results and Discussion}
\subsection{Comparison of \(\boldsymbol{H_c(x)}\) at low temperatures}
\begin{figure}
\includegraphics{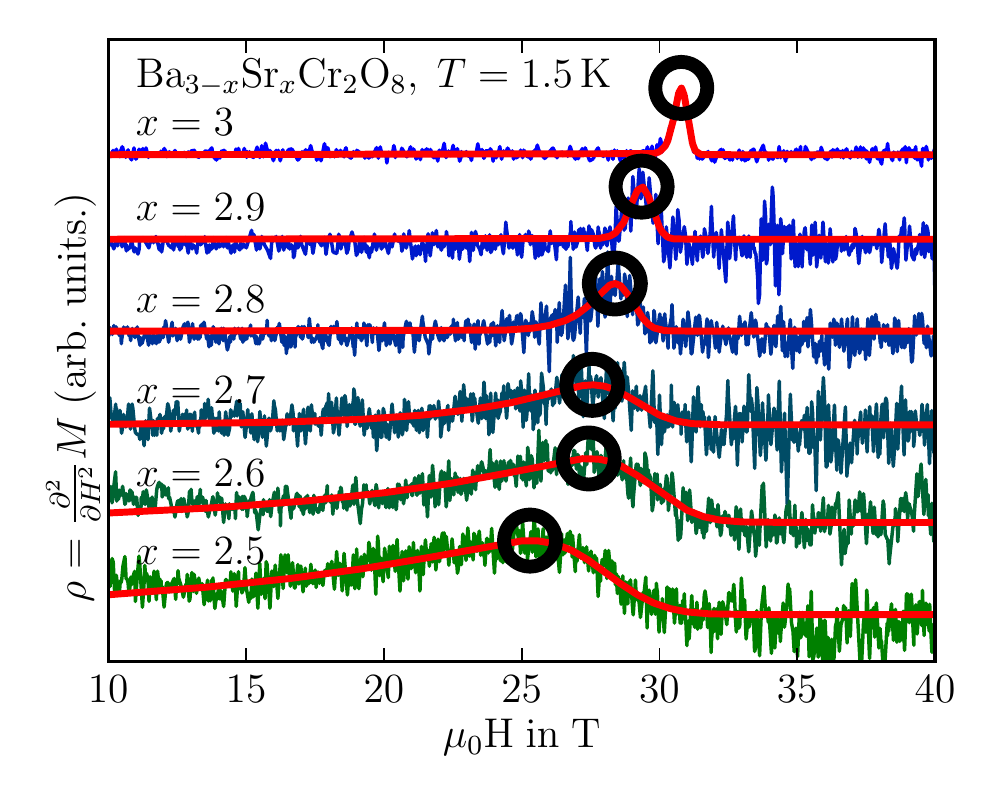}
\caption{Comparison of the second derivative of the magnetization \(\rho(H)\) as function of the applied magnetic field \(H\) for several different compositions of Ba\(_{3-x}\)Sr\(_x\)Cr\(_2\)O\(_8\). The data have been obtained from pulsed field magnetometry experiments at \(T\approx1.5\,K\). The red lines are fits according to Eq. \ref{eqn:fitfkt}, while the black circles correspond to the extrema of \(\rho\), marking the position of the critical field \(H_c\).}
\label{fig:comparison_at_low_T}
\end{figure}

The first important result of our analysis is a change of the critical field \(H_c\) as a function of the Sr content. In Fig. \ref{fig:comparison_at_low_T}, we have plotted \(\rho(H)\) for \(x\in\{3,2.9,2.8,2.7,2.6,2.5\}\) as obtained from magnetometry experiments in pulsed fields up to 60\,T at \(T\approx1.5\,\)K. For all examined values of \(x\), \(\rho\) can be well described using Eq. \ref{eqn:fitfkt}. The maximum of this function shifts towards lower magnetic fields for intermediate values of the Sr concentration \(x\), indicating a lowering of the critical field of the triplon condensation. In addition, the width of the peak increases drastically for intermediate stoichiometries, indicating a significant broadening of the transition. A similar, but less pronounced broadening of the antiferromagnetic transition has been be found in the solid solution of ungapped antiferromagnetic materials like jarosite\cite{nocera_jarosite}. In the case of Ba\(_{3-x}\)Sr\(_x\)Cr\(_2\)O\(_8\), we attribute this broadening to the increasing disorder that has been reported for this system.\cite{grundmann_ww-aenderung}

\begin{figure}
 \includegraphics{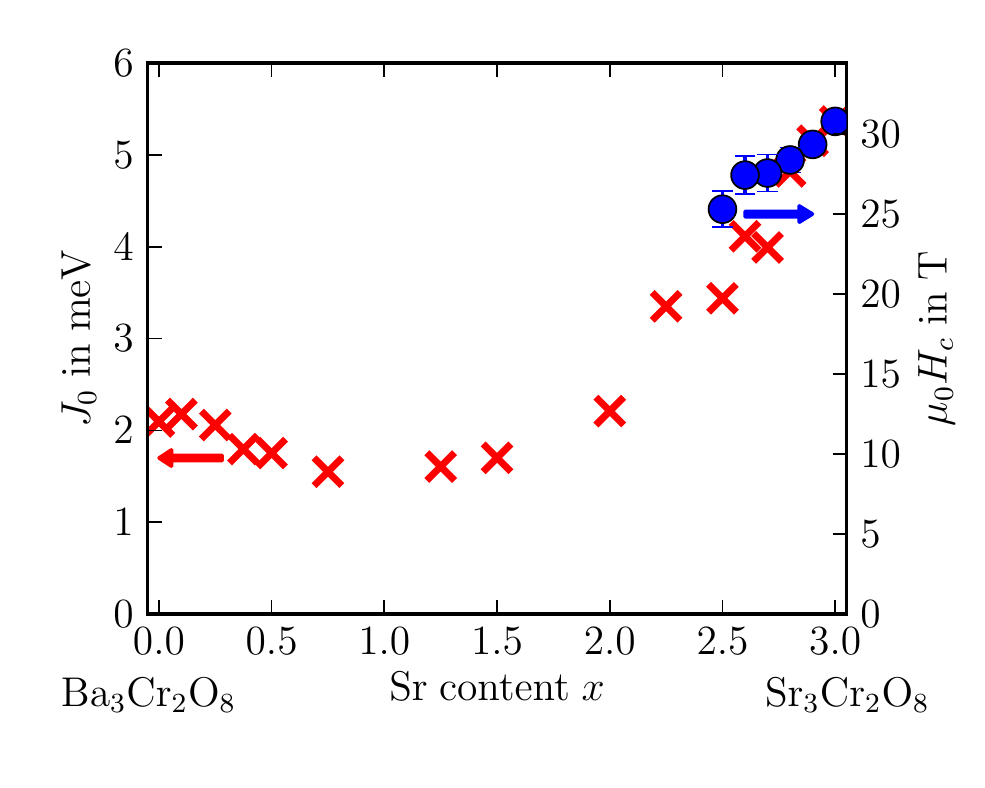}
\caption{Comparison between the interaction constant \(J_0\) from Ref. \onlinecite{mrbpaper} and the critical field \(H_c\) at \(T\approx1.5\,K\) as a function of the Sr content \(x\). The axes have been chosen in such a way that \(J_0(x=3)\) and \(H_c(x=3)\) coincide.}
\label{fig:comparison_BkritandJofx}
\end{figure}

As described above, a change of the critical field can be induced by a corresponding variation of the magnetic interactions in the system. Such a variation of the intradimer interaction constant \(J_0\) has been reported for Ba\(_{3-x}\)Sr\(_x\)Cr\(_2\)O\(_8\) due the partial replacement of Ba by Sr.\cite{mrbpaper} In Fig. \ref{fig:comparison_BkritandJofx}, we compare this change with the variation of \(H_c\) as a function of \(x\). As the trends for \(J_0(x)\) and \(H_c(x)\) coincide well for all examined values of \(x\), we conclude that the observed decrease of the critical fields at low temperatures can probably mainly be attributed to respective changes of the magnetic interactions.

\subsection{Changes to the phase boundary \(\boldsymbol{T_c(x,H)}\)}
\label{sec:phasengrenze}

\begin{figure}
\includegraphics{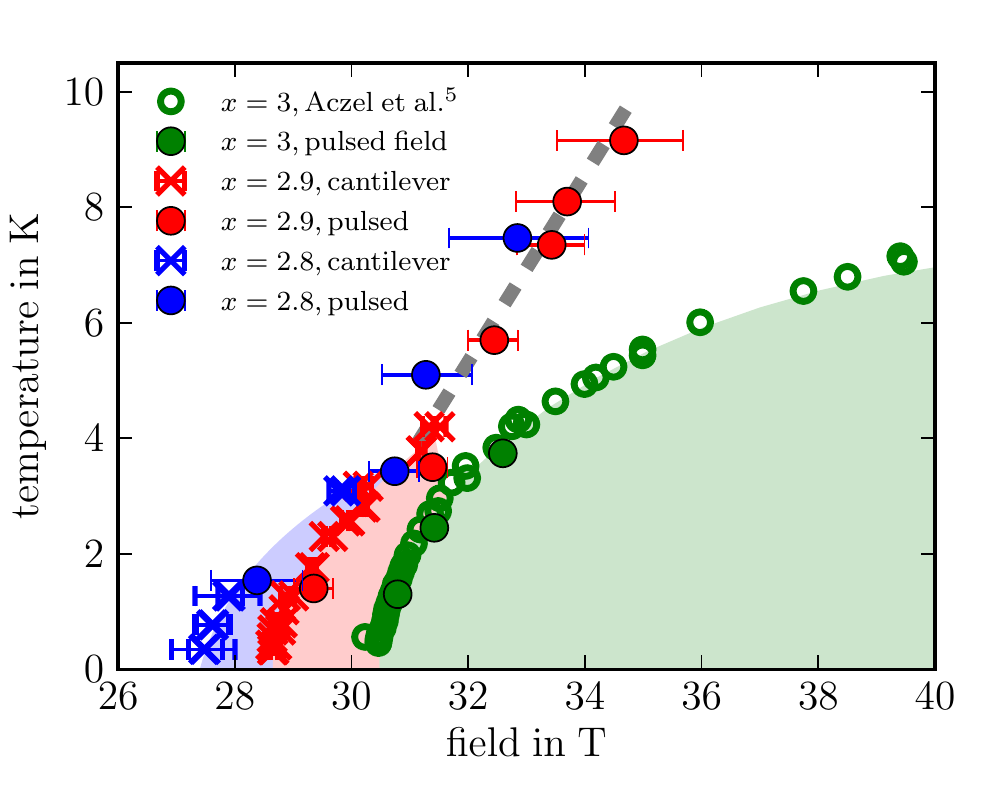}
\caption{Comparison of the critical temperatures \(T_c\) of the triplon condensation, as functions of the applied magnetic field for \(x\in\{3,2.9,2.8\}\). The open circles symbolize data points taken from Ref. \onlinecite{aczel_sr3cr2o8}. The solid colored areas and the dashed line are to guide the eye.}
\label{fig:phasendiagramm}
\end{figure}

The simplest possible change of the phase boundary \(T_c(H)\) with varying \(x\) would be a shift of the whole "dome" towards lower magnetic fields near the quantum critical point. However, our data show that this scenario does not hold for Ba\(_{3-x}\)Sr\(_x\)Cr\(_2\)O\(_8\) and the changes to the phase boundary due to a change of the Sr content go beyond a simple shift towards lower magnetic fields. In Fig. \ref{fig:phasendiagramm}, we have plotted the critical temperatures as functions of the magnetic field for Ba\(_{0.2}\)Sr\(_{2.8}\)Cr\(_2\)O\(_8\), Ba\(_{0.1}\)Sr\(_{2.9}\)Cr\(_2\)O\(_8\) and pure Sr\(_3\)Cr\(_2\)O\(_8\). The obtained phase boundaries do show a clear shift towards lower critical fields for temperatures below \(T\approx3.5\,\mathrm{K}\), in accordance with the reported change of the magnetic interaction constant \(J_0\).\cite{mrbpaper} It should be noted, that besides a decrease of the intradimer interaction \(J_0\), any change of the interdimer interactions \(J'\) would also alter the triplon band, and change both the lower and upper critical fields accordingly. At present, no information about the upper critical field or about a change of the interdimer interactions in Ba\(_{3-x}\)Sr\(_x\)Cr\(_2\)O\(_8\) as a function of \(x\) is available. We note that the domes \(T_c(H)\) for \(x=2.9\) and \(x=2.8\) appear to be somewhat flatter than that for \(x=3.0\) (see Fig. \ref{fig:phasendiagramm}), which may be a consequence of a certain dependence of \(J'\) on \(x\) or of the presence of disorder.

Changes in the magnetic phase diagram of quantum magnets due to variations in the chemical composition have indeed been explained based on various scenarios. A partial substitution of Ba by Sr very probably leads to structural disorder\cite{grundmann_ww-aenderung} or even free defect spins\cite{mrbpaper} in the system. Such a chemical disorder has been suggested to lead to a Bose-Glass transition in Tl\(_{1-x}\)K\(_x\)CuCl\(_3\)\cite{yamada_boseglass} and NiCl\(_{2-x}\)Br\(_x\)\(\cdot\)4SC(NH\(_2\))\(_2\) \cite{yu_dtn} (Br-DTN), thereby reducing the critical field and changing the critical exponent. Such a reduction of \(H_c\) is, as in the case of Br-DTN, often a simple consequence of a reduced spin gap\cite{zheludevdtn2015}. For Ba\(_{3-x}\)Sr\(_x\)Cr\(_2\)O\(_8\) we cannot make such a comparison as detailed information about the dispersion relation of the triplons and thus a direct measure of the spin gap is not available at present.
One of the most important features of a Bose-Glass is the finite compressibility, which translates to a finite magnetic susceptibility for \(T\rightarrow0\) and \(H<H_c\). The paramagnetic contributions due to free spins is too large in our samples to draw any decisive conclusion regarding the low field susceptibility of our samples, however. In any case, the magnetic phase boundary for a Bose-Glass of triplons should be \emph{tangential} to the field axis for \(T\rightarrow 0\),\cite{fisher, yamada_boseglass} rather than perpendicular to it for a three-dimensional triplon BEC. Our data do not indicate such a behavior although measurements at even lower temperatures would be necessary for absolute confidence.

%
%

For temperatures above \(T\approx3.5\,\mathrm{K}\), our experimental \(\rho(H)\) data show peaks that do not fit to the expected dome-like shape of the phase boundary for the triplon-BEC transition. These features, marked by a straight dashed line in Fig. \ref{fig:phasendiagramm}, are reminiscent of a phase boundary or crossover other than the appearance of a long range \(XY\)-order of the spin system. Such a behavior of \(M(H)\) has also been found in single crystals of pure Sr\(_3\)Cr\(_2\)O\(_8\).\cite{aczel_phd} It has been suggested \cite{Zapf} that the closing of the spin gap can lead to similar features, e.g. in the \(T(H)\) traces when measuring the magnetocaloric effect of a gapped spin system. 

\begin{figure}
\includegraphics{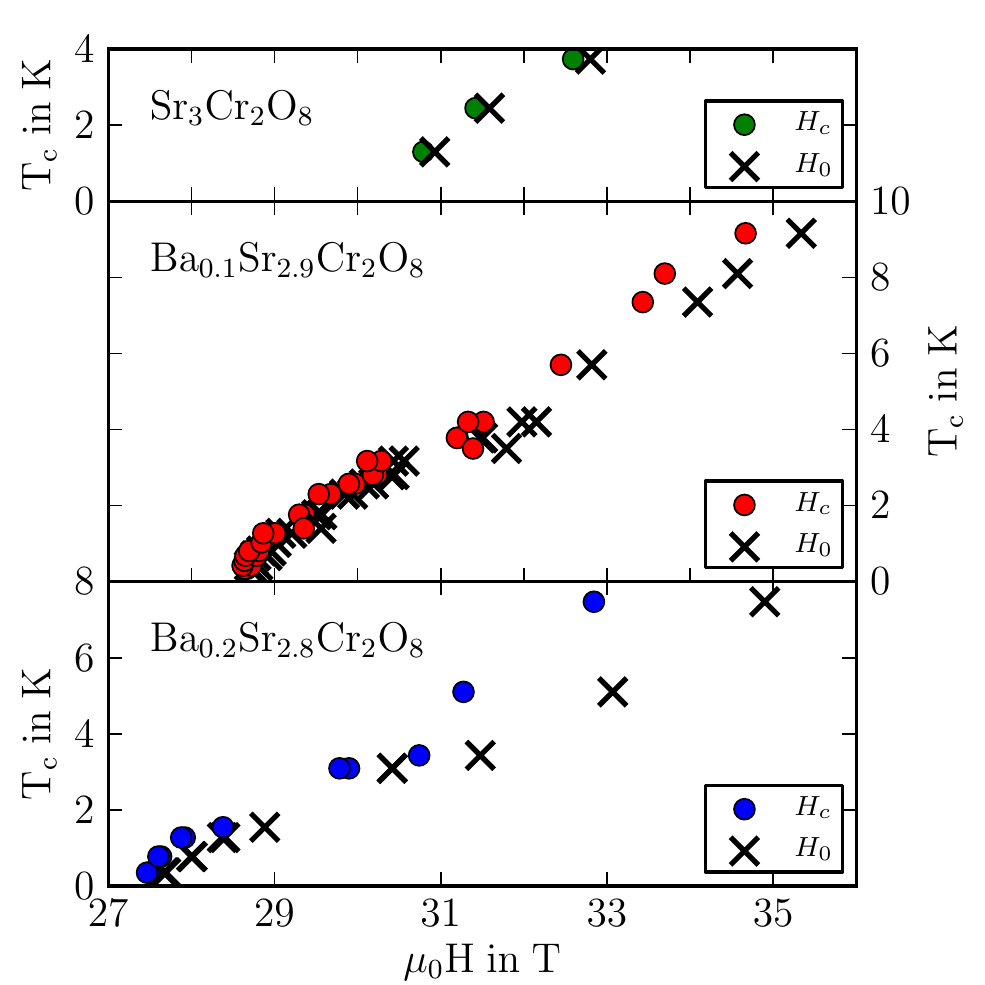}
\caption{Comparison of the values for the critical field \(H_c\) obtained as the maximum of \(\rho(x)\) (colored circles) and as the fitting parameter \(H_0\) in Eq. \ref{eqn:fitfkt} (black crosses) for Sr\(_3\)Cr\(_2\)O\(_8\), Ba\(_{0.1}\)Sr\(_{2.9}\)Cr\(_2\)O\(_8\) and Ba\(_{0.2}\)Sr\(_{2.8}\)Cr\(_2\)O\(_8\) at different temperatures.}
\label{fig:twokritfields}
\end{figure}

We would like to point out that determining \(H_c\) from the maximum of the observed \(\rho(H)\) of our samples is the common choice, but it may be not the only one. As described above, our data could be well fitted using Eq. \ref{eqn:fitfkt}. The term
\begin{equation*}
\frac{\mu_0}{\sqrt{2\pi}\sigma}e^{-\left(\frac{\mu_0(H_c-H_0)}{\sqrt{2}\sigma}\right)^2} 
\end{equation*}
can be interpreted as a Gaussian distribution of critical fields \(H_c\) around an average critical field \(H_0\) with variance \(\sigma^2\) that is given by the slight anisotropy\cite{kofu_ba3_bec} of the Landé g-value and local strain and crystal imperfections.\cite{zheludev} This distribution is then convoluted with an exponential decay 
\begin{equation*}
e^{\frac{\mu_0(H-H_c)}{\alpha k_BT}}\theta(H_c-H) 
\end{equation*}
towards low magnetic fields. The fitting parameter \(H_0\)  gives the center of the Gaussian distribution. For a purely Gaussian fitting function, \(H_0\) would thus be (and usually is) regarded as the \emph{actual} critical field of the triplon condensation. Concordantly, we have plotted the critical field \(H_c(T)\) as obtained from the maximum of \(\rho(x)\) in comparison to \(H_0(T)\) in Fig. \ref{fig:twokritfields}.

For all examined samples and temperatures, \(H_0\) is larger than \(H_c\). However, for Sr\(_3\)Cr\(_2\)O\(_8\) and based on the literature values for the critical field for the triplon BEC, this difference is too small to decide whether \(H_c\) or \(H_0\) should be regarded as the \emph{actual} critical field. The difference increases for higher temperatures and smaller values of the Sr content \(x\). However, both \(H_c\) and \(H_0\) decrease for decreasing values of \(x\) in a similar fashion. Thus, the main result of a shifted phase boundary towards lower magnetic fields with decreasing \(x\) does not depend on whether \(H_c\) or \(H_0\) is regarded as the \emph{actual} critical field.

\section{Summary}
We have prepared polycrystalline samples of Ba\(_{3-x}\)Sr\(_x\)Cr\(_2\)O\(_8\) based on standard solid state reaction schemes. Using pulsed field magnetometry and cantilever magnetometry experiments at temperatures down to \(340\,\mathrm{mK}\), we observe a decrease of the critical field with decreasing Sr content \(x\). This decrease is in accordance with the reported change of the magnetic interaction constant \(J_0\) as a function of \(x\). The observed changes in the magnetic phase diagram upon partial substitution of Sr by Ba may also be influenced by the presence of disorder, which will require further investigations.

\section{Ackowledgements}
We thank Erik Wulf for stimulating discussions. We acknowledge the support of the HLD at HZDR, member of the European Magnetic Field Laboratory (EMFL). We also acknowledge the support of the HFML-RU/FOM, member of the European Magnetic Field Laboratory (EMFL). This work is part of the research programme of the \emph{Stichting voor Fundamenteel Onderzoek der Materie (FOM)}, which is financially supported by the \emph{Nederlandse Organisatie voor Wetenschappelijk Onderzoek (NWO)}. This work was supported by the Swiss National Science Foundation Grants No. 21-126411 and 21-140465.
\bibliography{ba3-xsrxcr2o8_kritfeld}

\end{document}